\documentclass[11pt,a4paper]{iopart}
\usepackage{graphicx}
\usepackage{amssymb}
\usepackage[numbers]{natbib}
\newcommand{\newblock}{\em}

\begin{document}

\title[Granular Segregation]{Granular segregation in dense systems: the role of statistical mechanics and entropy}

\author{Matthias Schr\"oter$^1$ and Karen E. Daniels$^{1,2}$}

\address{$^1$Max Planck Institute for Dynamics and Self-Organization (MPIDS), 37077 G\"ottingen, Germany}
\address{$^2$Department of Physics, North Carolina State University, Raleigh, NC, 27695, USA}

\begin{abstract}
Granular segregation is ubiquitous in industrial, geological or daily-life context, but there is still no unifying
theoretical approach. In this review, we examine two examples of granular segregation 
-- shallow rapid flows and rotating drums -- which suggest
that the dynamics of systems at intermediate and high density
might be amenable to a statistical mechanics approach. 
\end{abstract}

\maketitle

\begin{quote}
 
{\itshape He wondered how long it would take before everything in the bowl was completely mixed. But what did that mean, completely mixed? Every grain of the ingredients would have to be distributed perfectly, the particles of salt and baking soda spaced just so throughout the flour, each one a fixed distance from all the rest. He tried to picture it, a solid, three-dimensional white field supporting a dense and uniform lattice of particles of other shades of white. And what about the flour itself, no two grains of which were alike---how could that be completely mixed, even if there were no other ingredients present in it, making their own pattern? And how would he know when that moment of perfect distribution had been achieved?}

\hfill  John Banville, {\itshape The Infinities}
\end{quote} 

\bigskip

As even fictional characters have come to notice, the mixing of granular materials is a tricky business. 
The very processes by which you might hope to evenly distribute particles of different type -- stirring or shaking -- 
quite commonly, in fact, create domains of different particle types \citep{Ottino2000, Kudrolli2004}.
The aim of this review is to explore the problem of mixing and segregation in dense granular systems 
within the context of statistical mechanics, and then to use this approach to highlight open research questions. 

We begin by reviewing why ordinary fluids either mix or phase-separate, and then discuss the extent to which granular materials exhibit similar and/or different behaviors. A familiar example is the success of kinetic theory in describing Soret-like effects in granular gases.  We follow with two examples from higher-density systems which have been more challenging to understand.
First, we examine segregation dynamics in shallow flows, where recent theories about the local
distribution of particles and voids \citep{Aste-2008-EGD, Clusel2009} provide hope for new approaches.
Second, we suggest that both axial segregation and its subsequent coarsening observed in rotating drums can
be understood as a surface tension driven instability within the statistical mechanics framework suggested by 
Edwards and coworkers \citep{Edwards1989, Mehta1989}.

\section{Segregation: ordinary vs. granular fluids}

Classical statistical mechanics -- in the form of thermal fluctuations, entropy, free energy, and gradients -- 
 has allowed us to understand why and how a mixture of oil and water will spontaneously phase-separate, while a 
mixture of vinegar and water will not. 
In {\it equilibrium thermal systems} (at fixed temperature and volume), the final state is governed by the minimization of free energy ${\cal F} =  U - TS $, and thermal fluctuations allow the system to explore many different configurations until the minimum is reached.  
For conventional fluids, the molecular sizes are typically similar enough that depletion effects 
are small; in such cases, the entropy $S$ will favor a mixed state. 
Therefore, the presence or absence of phase separation will largely be a consequence of minimizing the potential energy $U$. 
If we consider two species $1$ and $2$, the three contributions to $U$ are the interaction energies $U_{11}$, $U_{22}$, and $U_{12}$.
If  the interaction between unlike molecules, $U_{12}$, is less negative (smaller magnitude) 
than  $U_{11}$ and $U_{22}$, then the potential energy contribution will favor phase separation. 
If $\Delta U =  U_{12} - (U_{11} + U_{22})/2$ is large enough
to overcome the mixing entropy, the equilibrium state is a segregated one. 

Even if the system should, by these arguments, de-mix, the surface tension $\sigma$ between phases $1$ and $2$ can nonetheless cause the segregation process to be kinetically hindered. A droplet of size $R$ containing a single species loses free energy via the creation of an interface of surface area $A$, by an amount $\Delta {\cal F} = - A \sigma \propto R^2$. Simultaneously, it gains free energy proportional to its volume $V$, in an amount  $\Delta {\cal F} = V \, \Delta U \propto R^3$.  A nucleated droplet above a certain critical size $R_c$ will have a net negative change in free energy, and will therefore continue to grow in size; droplets smaller than $R_c$ will shrink. Therefore, a phase-separated state can only occur when thermal fluctuations  provide at least one nucleated droplet above this critical size. In the absence of such a droplet, the system can stay in the metastable mixed state indefinitely.

\begin{figure}[b]
\centerline{\includegraphics[width=5in]{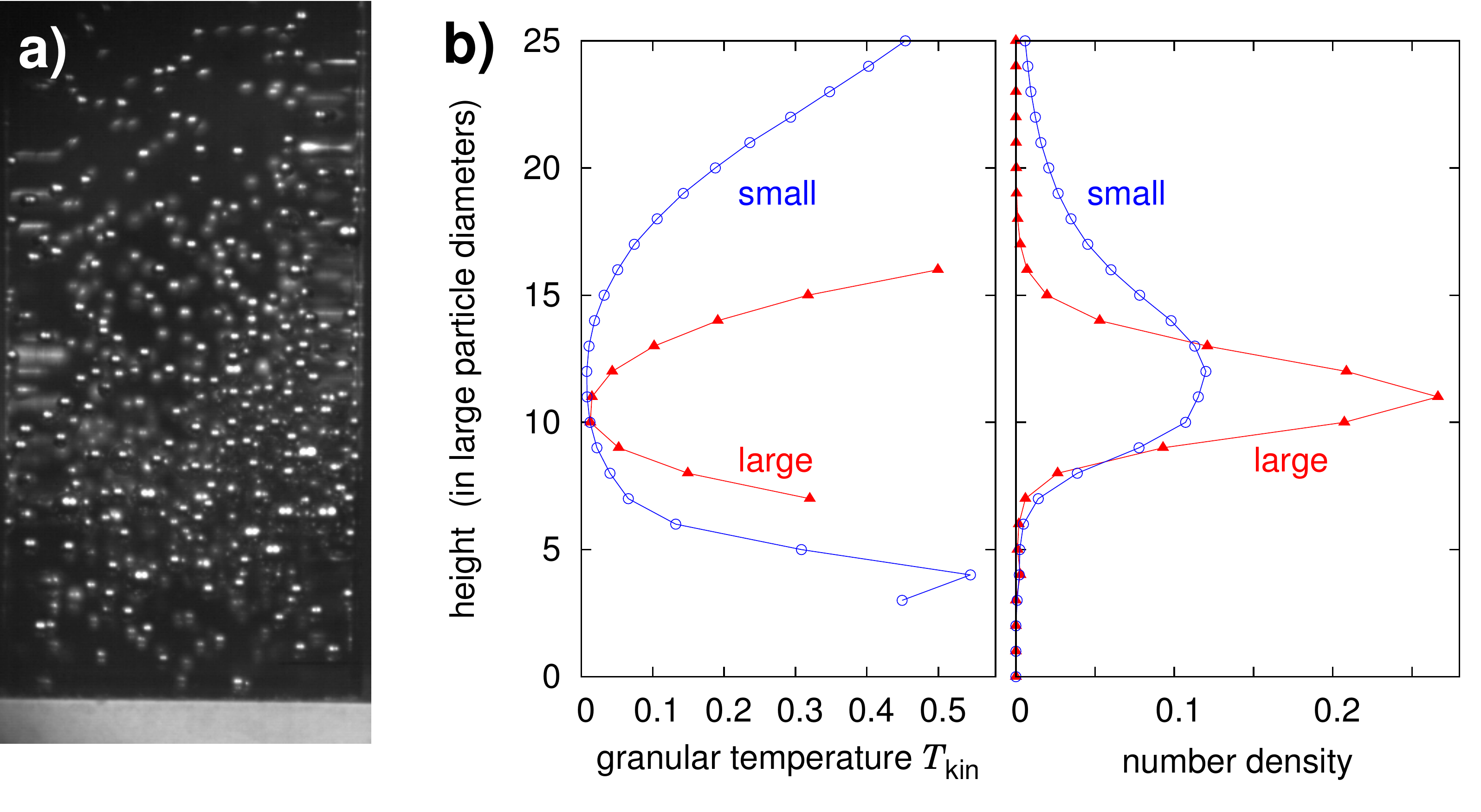}}
\caption{Thermal diffusion in a granular gas. (a)  Image of a vertically shaken, maximally expanded mixture of equal volumes of brass
spheres with 1.19~mm and 2.38~mm diameter.  Shaking parameters are 15 Hz and an average acceleration of $7 g$.
The larger particles accumulate in a small region in the lower half of the column.
(b) Results from a MD simulation which matches the experiments shown in panel a. The granular temperature
shows a pronounced minimum in which the large particles  preferentially accumulate. This agrees with the prediction of
granular kinetic theory  that the larger particles have a higher thermal diffusion coefficient.  Modified from \citep{Schroter2006}.
}
\label{fig:granSoret}
\end{figure}

When a thermal system is driven {\it out of equilibrium}, whether by gravitational, thermal, or other gradients, differences in the transport coefficients can provide additional sources of demixing. A notable example is the effect of thermal gradients on mixtures of ordinary fluids. A phenomenon known as thermophoresis \citep{Platten2006} breaks the symmetry such that molecules of differing masses migrate either up or down the thermal gradient. Typically, the heavier species movies towards the lower-temperature region, commonly known as the Soret effect. While thermophoresis is exploited in industrial processes, and thought to be an important factor in transport across biological membranes, the mechanisms underlying this behavior remain a matter of some debate \citep{Duhr2006}. In practice, the Soret coefficient remains an empirically-measured transport coefficient.

It is not a priori clear how many of these explanations will survive in the case of {\it granular systems}, which are formed out of
macroscopic constituents. 
First, the large size of the particles means that the thermal energy $k_B T$ is typically a 
factor of $10^{12}$ smaller than other energy scales. As a consequence, granular systems are said to be {\itshape athermal}, 
and any dynamics are the consequence of external driving rather than thermal fluctuations. 
Second, due to both friction and inelastic collisions, 
the system is dissipative: a steady energy influx is required for any persistent dynamics, and hence any
mixing or segregation of particles.

In spite of these caveats, there is hope that a statistical mechanics-like approach can elucidate granular dynamics. 
In the case of dilute granular gases, segregation can be understood on the basis of an appropriately-extended kinetic theory.
 This approach first
 defines a granular temperature $T_g$ as the averaged kinetic energy of the random motion of the particles
\citep{Brilliantov2004}.
Within this framework, it is possible to compute the dependence of the transport coefficients on the particle properties. 
Because most granular systems are driven by injecting energy at the boundaries, gradients in $T_g$ are ubiquitous.
 An example is shown in Fig.~\ref{fig:granSoret}, in which the  large particles accumulate
at a well-defined temperature minimum. Such Soret effects have been well-described by granular kinetic theory
\citep{hsiau:96,jenkins:02,Galvin:05,brey:05,Serero2006,garzo:06,Garzo2008,garzo:11,brey:11}.

\section{Vertical segregation in rapid, shallow flows} 

There is a common intermediate regime of granular flows, where particles are freely-flowing but nonetheless experience significant constraints on their motion due to the nearness of neighbors. In such cases, the assumptions of kinetic theory have broken down. Such flows are easily found in both natural systems (debris and pyroclastic flows \citep{Iverson2001}) and many industrial processes \citep{Bridgwater1976}.

Such flows are commonly shallow, and large particles segregate to the free upper surface of the flow. Therefore, theoretical approaches have paid particular attention to this geometry. Models by a number of groups \citep[][e.g.]{Savage1988, Dolgunin1995, Gray2005, Marks2011a} rely on the idea of {\itshape kinetic sieving} as the source of size-segregation. In each case, the underlying cause lies in the comfortable and oft-stated idea that as sheared layers flow past each other, small particles are more likely to find voids to fall into than are the large particles \citep{Rosato1987}. This imbalance in the downward flux of small particles as compared to large particles, combined with a continuity equation for the conservation of mass, will generically lead to size-segregation. However, this sieving effect has yet to be derived from first principles. 

\citet{Savage1988} provided an early attempt at a quantitative model, which they based on a very general maximum-entropy approach. In this approach, a particle has a probability ${\cal P}(\varepsilon)$ of finding a void of lateral size $\varepsilon$
\begin{equation}
 {\cal P}(\varepsilon) = \frac{1}{{\bar \varepsilon} - \varepsilon_0}
\exp \left[ - \frac{\varepsilon - \varepsilon_0}{{\bar\varepsilon - \varepsilon_0}} \right]
\label{e:SavLunVoid}
\end{equation}
with ${\bar \varepsilon}$ the mean void size and $\varepsilon_0$ the minimum void size. This probability is then used to calculate the probability of one layer of the flow capturing a particle from a neighboring layer. The resulting model provides a prediction for the steady-state particle size distribution in a steady shear flow.

\begin{figure}
\centerline{\includegraphics[width=5in]{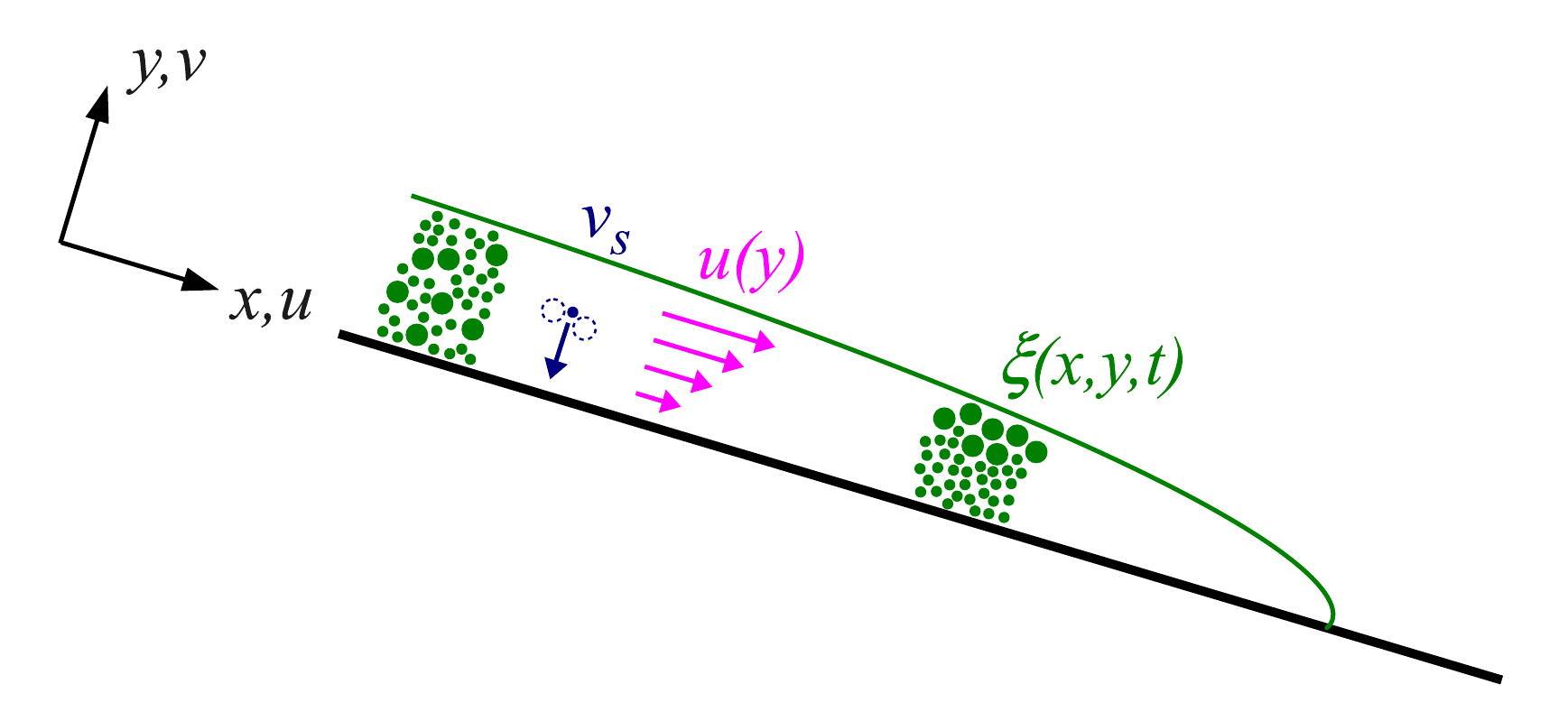}}
\caption{Schematic drawing of segregation within an shallow flow. An initial mixture with concentration $\xi =  \frac{1}{2}$ (at all depths) evolves to one in which $\xi \approx 0$ at the upper surface and $\xi \approx 1$ at the base of the flow, after exposure to a flow velocity $\vec v = (u,v)$ which generates a segregation velocity $v_s$ via kinetic sieving.}
 \label{f:avalanche}
\end{figure}

One of the shortcomings of the \citeauthor{Savage1988} model is that it predicts segregation even in the absence of gravity. The model of \citeauthor{Gray2005} solved this difficulty by introducing momentum-balance in deriving an equation for the evolving concentration field $\xi(x,y,t)$ of the small particles, where $\xi = 1 \, (0)$ corresponds to all small (large) particles. For the situation shown in Fig.~\ref{f:avalanche}, 
\begin{equation}
\frac{\partial}{\partial t} \left[ \xi   \right] +
\frac{\partial}{\partial x} \left[ \xi u \right] + 
\frac{\partial}{\partial y} \left[ \xi v \right] 
  - \frac{\partial}{\partial y} \left[\xi  \, v_s(\xi) \right]
=0,
\label{e:Gray}
\end{equation}
where ${\vec v} = (u,v)$ is the velocity field of the bulk granular material in the ${\vec x} = (x,y)$ directions, as determined from an either experimental measurements or a rheological model. A {\itshape segregation velocity} $v_s(\xi)$ is superimposed on the small particles, relative to the bulk motion ${\vec v}$.
In \citet{Gray2005}, the functional form is taken to be $v_s(\xi) = S (1-\xi)$, so that the last term of Eq.~\ref{e:Gray} is nonlinear. This is a lowest-order (linear perturbation) model in which the large particles (which have a concentration $1-\xi$) experience a pressure-imbalance with respect to the small particles, and $S$ is the non-dimensional strength of the segregation velocity.
While the model has recently seen some semi-quantitative success in predicting the dynamics of both experiments and DEM simulations, \citep{May-2010-SDS, Wiederseiner2011}, the correct functional form of $v_s(\xi,\phi, P, \cdots)$ remains little-investigated. 

In particular, it remains unclear how to correctly incorporate the partial and/or global pressure $P$.
Contrary to the assumptions of the model, \citet{Smart2008} used simulations to observe that over a large parameter regime, the pressure on small and large particles is in fact close to equal. Furthermore, \citet{Golick2009} observed that the rate at which shear-segregation occurs (related to $S$) is a non-monotonic function of the particle size ratio, with dissimilar sizes experiencing the most sensitivity to the global pressure. 
Finally, in this Focus Issue, \citet{Fan2011a} have had success with a model similar to that of Gray and Thornton, with the addition of terms which account for gradients in kinetic stresses, as measured through velocity fluctuations. 

Modern experimental techniques, as well as discrete element simulations, now allow us to measure both the local void distribution and the 
trajectory of each particle. This provides all of the local dynamics needed to directly observe particles falling into voids. 
Fundamentally, the assumptions behind the kinetic sieving mechanisms are both empirically testable 
(``What is the difference in pressure on large vs. small particles?'') and related to recent theories 
about the statistical mechanics of static granular materials. Using theories which describe the distribution of 
free space \citep{Aste-2008-EGD, Clusel2009} 
it should be possible to calculate transition probabilities for either large or small grains within a bath of a
 distribution of known void sizes. Furthermore, theories which include the stresses on the grains 
\citep{Ball2002, Henkes-2009-SMF, Tighe2011a}
 would permit the direct calculation of pressure-imbalances. 

\begin{figure}
\includegraphics[height=2.8in]{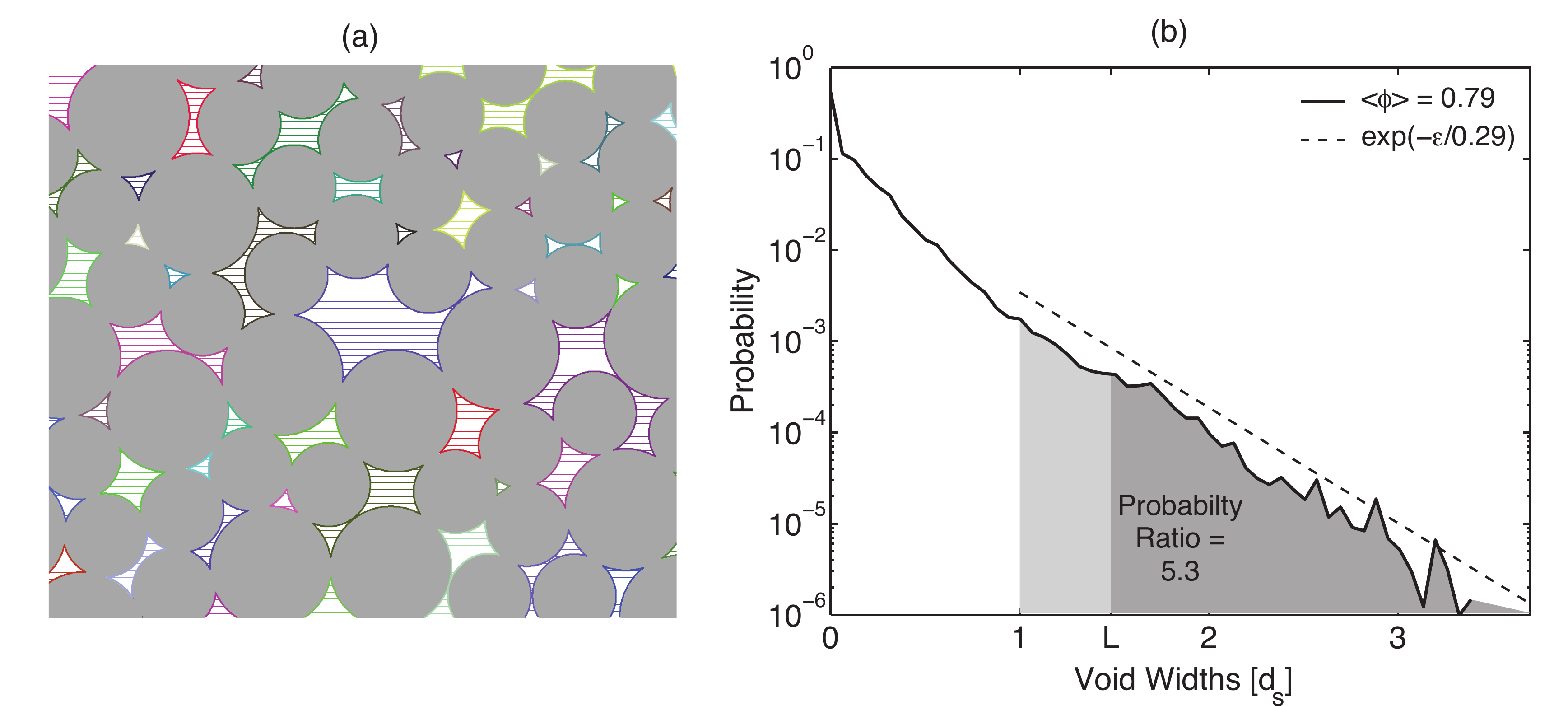} 
\caption{
(a) Sample image showing detected voids (local $\varepsilon$), using previously-published dataset \citep{Puckett-2011-LOV} of an agitated two-dimensional floating on a horizontal air table. The widths of the voids are measured in the direction shown by the colored lines. Each color represents a separately-detected void. 
(b) The distribution of measured void widths at $\phi \approx 0.8$. Dashed line is a fit to an exponential distribution. 
}
\label{f:voiddist}
\end{figure}

Interestingly, ensemble-approaches within the granular physics community have focused on the local $\phi$ distributions, whereby all un-occupied space is partitioned up and associated with the nearest particle. Two popular measures of $\phi$ are the radical Vorono\"i tessellation and the navigation map \citep{Richard2001}. However, other scientific communities \citep{Oda1972, Oda1982, Stoyan2011, Ghedia2012} have focused on the void-size ($\varepsilon$) distribution, an example of which is shown in Fig.~\ref{f:voiddist}a for a two-dimensional granular system. The void distribution is of primary importance for segregation problems, as a means to construct an empirical or analytical expression of the maximum-entropy ideas first laid out by \citet{Savage1988}. In Fig.~\ref{f:voiddist}b, the approximately exponential shape is similar to what is predicted by Eq.~\ref{e:SavLunVoid}. The light/dark shaded regions represent the probability of finding a void space which is large enough for small/large particles to move into. The ratio of these two probabilitiess, even for the simlarly-sized particles shown here, is more than 5. Direct measurements of $v_s$ as a function of the local environment $(\xi, P, \phi, \varepsilon)$ will allow us to determine the most relevant dependencies.

\section{Axial segregation and coarsening in a rotating drum}

In many industrially- or geophysically- relevant systems, the flow of particles is less rapid and collisional than those 
described in the previous section, and particles are instead densely-packed and in sustained contact with each other. 
Rather than strong gradients driving the flow, particles slowly roll and slide past each other. The paradigmatic case for 
segregation in this dense-flow regime is the rotating drum: a partially-filled cylindrical container containing two sizes 
of grains which rotates slowly around its horizontally-oriented  axis \citep{aranson:06,seiden:11}.

When the rotating drum is tube-shaped (longer than it is thick), three sequential segregation stages can be distinguished. 
Within the first few rotations, {\it radial segregation} occurs: the small particles migrate towards the central axis of the drum and form a 
channel extending the whole length of the cylinder. This channel is
surrounded by a nearly-pure layer of large particles; it forms via a kinetic sieving similar to that discussed in the previous section.

\begin{figure}
\centerline{\includegraphics[width=6in]{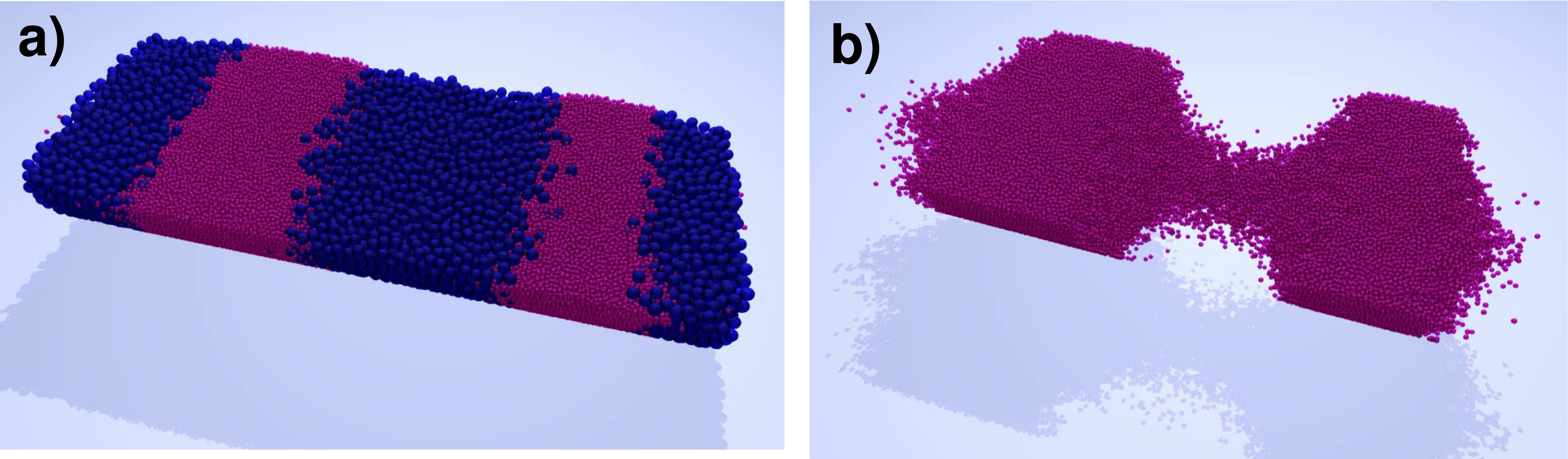}}
\caption{Axial segregation pattern in a rotating drum experiment.
The tube was half filled with equal volumes of particles with diameters 
1~mm and 0.425~mm, and then rotated at 15~rpm 
until the axial stripes had formed. Image (a) shows the outer appearance of the bed, while (b) highlights
the connecting core between the same two stripes, by removing the large particles from the image. 
Both images are rendered with POV-Ray using experimental particle positions determined by X-ray tomography.
A more complete visualization is provided by the accompanying Movie 1.
}
\label{fig:axial_segregation}
\end{figure}

During the next ten to hundred rotations, a process called {\it axial segregation} happens: the initial core of small particles becomes 
unstable to longitudinal modulations and finally  alternating bands of 
large and small particles appear along the axis of the drum. In addition, the bands of small particles remain connected by the radially-segregated 
 core of small particles as shown in Fig.~\ref{fig:axial_segregation}; see also the accompanying Movie 1.

The basic mechanism behind axial segregation is still an open question. As most of the grain motion happens at the free surface, 
it was initially thought that axial segregation is due to differences in the dynamic angle of repose of the two particle sizes \citep{zik:94}.
Indeed there are clear differences in the surface flow profiles of the two sizes \citep{taberlet:06}.
This idea has subsequently been expanded to a full continuum description by adding the relative concentration of particles 
\citep{aranson:99,aranson:99_PRE}. 

With the advent of three-dimensional imaging measurements (primarily MRI) it became evident that  
 subsurface flow (or, more precisely, the migration of small particles in the radial core)
 is the mechanism which creates the axial segregation pattern
\citep{hill:97,hill:97_PRL,khan:04,juarez:08,Nguyen2011}. 
This result has also been supported by both numerics \citep{taberlet:06,rapaport:07}
and experiments showing that mixtures of same size spheres but different angles of repose do not show 
any axial or radial segregation \citep{pohlman:06}. 
However, the question of the driving force behind the flow through the core is still not answered. While it has been shown that
frictional interactions at the end walls can support the formation of axial bands \citep{chen:11}, axial segregation 
appears simultaneously everywhere in long drums, rather than at first in the vicinity of the end walls \citep{fiedor:06,Finger2006,juarez:08}. 
Below, we suggest a new origin for this radial core flow. 

\begin{figure}
\centerline{\includegraphics[width=6in]{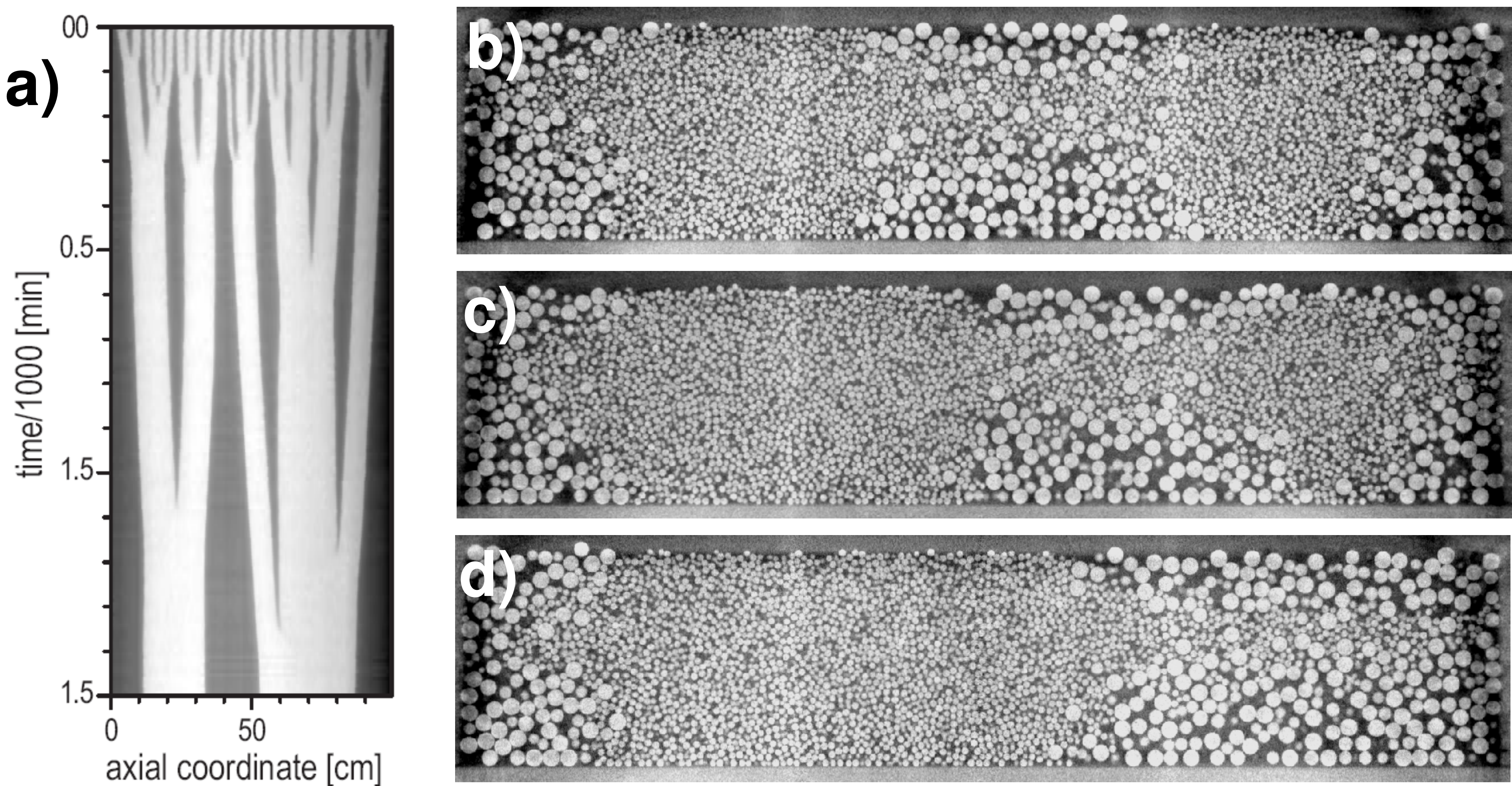}}
\caption{Coarsening dynamics in a rotating drum experiment. (a) Space-time plot of the evolution inside a long rotating drum. 
Dark areas correspond to bands of small glass spheres (0.55~mm diameter), bright areas consist of large  glass spheres (1.5~mm).
The drum rotates at 15~rpm. Reprint from \citet{Finger2006}.  
(b, c, d) Vertical cross section through the middle of a rotating drum  initially containing two bands of small particles.
Images are taken with x-ray tomography after 100, 9500, and 24300 rotations with 20~rpm. 
Courtesy Tilo Finger.
}
\label{fig:rot_drum}
\end{figure}

Finally, on timescales of hundreds to many thousands of rotations, these axially-segregated bands undergo {\it coarsening}. 
During this stage (see Fig.~\ref{fig:rot_drum}a), the large-particle bands grow together and merge, while the number of 
small-particle bands continually decreases. \citet{Finger2006} used carefully-prepared samples containing only two bands of
small particles to show that it is typically the band of smaller volume which vanishes. 
The process by which this happens corresponds to a continuous flow of particles through the connecting radial channel, 
rather than a diffusive random walk  \citep{arndt:05,taberlet:06,Finger2006,rapaport:07,Nguyen2011}.
The only known theoretical approach capable of explaining the coarsening dynamics is the 
continuum theory laid out in \citep{aranson:99,aranson:99_PRE}, but it ignores the dynamics inside the core.

In this review, we suggest that a unified explanation may be able to account for both the axial segregation 
and the subsequent coarsening. 
It is helpful to first consider the often-used analogy of a granular fluid \citep{Jaeger1996}: axial segregation
is morphologically similar to the
 Rayleigh-Plateau instability in ordinary fluids, whereby a cylindrical fluid column breaks
up into individual drops. The instability is driven by the reduction in surface energy provided by the smaller surface area of the droplet state.

In the case of granular fluids, such an explanation would depend on the existence of a effective surface tension $\sigma$
between the two fluids (small-particle and large-particle).\footnote{Such a surface tension is distinct from the effective surface tension 
which arises due to small cohesive forces, recently found in freely falling granular streams \citep{Royer2009,Waitukaitis2011}.}
Such a surface tension, to be described below, would also provide an
intuitive explanation for the coarsening dynamics: it is a consequence of 
the system continuing to reduce its interfacial area by merging several small-particle bands into a single, extended band. 
It is worth noting that the kinetic sieving mechanism discussed 
in the last section provides a stabilizing mechanism at the necks which connect the individual bands/droplets. 
Only because the necks don't rupture as in an ordinary fluid (see Fig.~\ref{fig:rot_drum}bcd), 
the small-particle bands can continue to exchange ``fluid'' till they have merged into one extended stripe.

How might such an interfacial free energy arise between two granular fluids? In dense granular system, the interparticle interactions 
take the form of Hertzian normal forces and frictional tangential forces. Neither of these forces depends 
strongly on particle size. Therefore, the free energy will mostly be determined by entropic contributions. 
The enduring or sliding contacts make the system quite different from a classical hard sphere systems, 
where the configurations with even only two particles touching are of measure zero. 
Therefore we will turn to the Edwards ensemble \citep{Edwards1989,Mehta1989} 
for a framework in which to discuss the relevant configurational entropies.

Edwards and coworkers suggested the utility of a configurational entropy
\begin{equation}
S_{conf} = \lambda \, \ln \Omega,
\end{equation}
where $\Omega$ is the number of all possible mechanically stable configurations 
under constant volume fraction and boundary stresses,  and 
$\lambda$ the as-yet-undetermined granular equivalent to the Boltzmann constant. 
While there are a number of open questions, for instance the role of ergodicity, a growing community has started to 
work on such a statistical mechanics approach based on the Edwards ensemble \citep{PicaCiamarra2012}. 

For the purposes of determining an interfacial tension, the salient observation is that the number of possible packings of monodisperse 
spheres does not depend on their diameter at given volume fraction. Therefore,  $S_{conf}/N $ (entropy per particle) is also independent of the diameter for a pure
granular fluid. However, at an interface between two sizes of particles, the number of mechanically stable configurations can be expected to depend 
on the diameter ratio. As the ratio between the two sizes deviates from unity, there will be extra void space at the interface which can not be effectively used to create packings. 
This is similar to granular interfaces with flat walls \citep{Jerkins2008a, Desmond2009}. 
This unusable void space will result in a reduced value for $S_{conf}$, as compared to the value for a single-component fluid, and will 
therefore provide an effective surface tension.

Such an effective surface tension would also explain why the narrower of two small-particles bands is typically the one which vanishes, as shown in Fig.~\ref{fig:rot_drum}. The narrower band has a less favorable surface-to-volume ratio, resulting in fewer accessible 
configurations as compared to the wider band. Therefore, particles within the connecting channel 
will find themselves more often in configurations further away from the center of the narrower band. This results in a net flux from the narrower to wider band.

This analogy to interfacial tension remains, at present, a plausibility argument rather than a proof of the existence 
of an effective surface tension. 
We see two potential ways of testing this hypothesis. First, one could manually prepare a 
radially-segregated system in which the initial cylindrical interface between the two fluids is as smooth as possible.
 A series of X-ray tomographs covering the first few rotations will permit measurements of the evolution of the interface. If the evolution is 
indeed driven by an effective surface energy, a Fourier-decomposition of the interface will reveal a characteristic 
range of exponentially growing modes. In the Rayleigh-Plateau scenario \citep{Eggers1997}  
the wavelength of the fastest growing mode is independent of the value of $\sigma$. However,
the growth rate should depend on $\sigma$, which in turn can be modified by changing the size ratio of the particles.  

Second, one could use capillary wave theory to test for the existence of an effective surface tension \citep{Hoyt2001, Vink2005}. 
This would require a sheared setup in which the large particles initially form a layer on top of the small particles. 
As this sequence of layers is stable to kinetic sieving, the externally-imposed shear
will  create small fluctuations of the interface, which can again be captured by x-ray tomography.
The mean square amplitude $\langle A^2 \rangle$  of the Fourier modes of the interface will depend on the wavenumber $k$ and the
interface sidelength $L$. In the presence of an effective surface tension, 
$\langle A^2 \rangle$  should scale like
\begin{equation}
\langle A^2 \rangle \, \sim \, \frac{1}{L^2 \, \sigma \, k^2}  \;,
\end{equation}
which can readily be tested.

\section{Conclusion} 

In this review, we have described segregation in two specific granular systems of medium to high density. In both cases,
an approach based on entropic or probabilistic arguments seems promising explanations of observed phenomena. 
We believe that this likely holds true for most such granular systems: the kinetic energy does not influence the dynamics strongly, nor is it likely that there are significant differences in contact forces  between different particle-species. Instead, volume exclusion and geometrical constraints
have a strong, yet mostly unexplored, influence on the number of available configurations (entropy). Even though the
myriad ways of external driving such systems are presumably far from ergodic, it is the accessible phase-space volume which determines 
the fate of the system.    

It is interesting to note that while the question of segregation was pivotal to Edwards and coworkers when suggesting
their statistical mechanics approach \citep{Edwards1989, Mehta1989}, it has only recently begun to see application. The work to date has primarily concerned simulated granular systems \citep{Nicodemi2002, Srebro2003, Tarzia2004, Tarzia2005}, where particle-scale measurements of velocities, free volumes, and forces have been possible. With the increasing availability of experimental methods such as photoelastic force measurements, MRI, and X-ray tomography, 
the time is ripe to move forward understanding the particle-scale origins of these eye-catching bulk phenomena.

\section*{Acknowledgments} 
We thank Sibylle N\"agle for creating Fig. 4 and Movie 1, Tilo Finger for providing the raw data pertaining to Figs. 4 and 5, and  
Klaus Kassner for helpful discussions. K.E.D was supported by NSF CAREER award DMR-0644743, and by an Alexander von Humboldt Fellowship while in G\"ottingen. 
K.E.D. and M.S. are grateful to the Aspen Center for Physics for a visit during which many fruitful discussions occurred.


\end{document}